\documentclass[aps,preprint,psfig]{revtex4}

\usepackage{graphicx}

\usepackage[latin1]{inputenc}
\begin{document}

\title{Generic theory of active polar gels: a paradigm for cytoskeletal dynamics}

\author{K. Kruse$^1$, J.F. Joanny$^2$, F. J\"ulicher$^1$, J. Prost$^{2,3}$, K. Sekimoto$^{2,4}$}
\affiliation{$^1$Max-Planck Institut f\"ur Physik komplexer Systeme, 
N\"othnitzerstr. 38, 01187 Dresden, Germany}
\affiliation{$^2$Physicochimie Curie (CNRS-UMR168), Institut Curie, Section 
de Recherche, 
26 rue d'Ulm 75248 Paris Cedex 05 France}
\affiliation{$^3$E.S.P.C.I, 10 rue Vauquelin, 75231 Paris Cedex 05, France}
\affiliation{$^4$LDFC Institut de physique, 3 rue de l'Universit\'e, 67084
Strasbourg Cedex, France}
\date{\today }

\pacs{87.17Jj, 82.70Gg, 82.35Gh}
\begin{abstract}
We develop a general theory for active viscoelastic materials made
of polar filaments. This theory is motivated by the dynamics of the 
cytoskeleton. 
The continuous consumption of a fuel generates a non equilibrium state
characterized by
the generation of flows and stresses. Our theory can 
be applied to
experiments in which cytoskeletal patterns are set in motion
by active processes such as those which are at work in cells.
\end{abstract}

\maketitle

\section{Introduction}

Molecular biology has scored an impressive number of success stories over the last forty 
years. It has provided us with a vast knowledge about molecules at work in living systems, 
about their involvement in specific biological functionalities and about the structure of their chemical networks \cite{albe02}. 
Yet one is still unable to take advantage of this knowledge for constructing 
a comprehensive description of cell behavior. Assuming that we had all desired molecular 
informations, the computer time required to describe meaningful cell behavior would be 
totally prohibitive. Meanwhile there is a clear need for a generic description of cells.

An alternative approach consists in identifying a reduced
number of key characteristics and construct a phenomenological description of a "simplified" but
relevant cell. Again this task is currently too complex at the scale of a whole cell, but it can be
envisioned for some of its constituents. A good example is provided by the description of
membranes: for length scales of the order of a few tens of nanometers up, it can be considered 
as a fluctuating  surface on which a few densities are distributed and through which a few other
densities permeate passively or actively \cite{lipo95,mann01}. 
The construction of the theory has been going on 
for more than thirty years. It sheds light on the physics of membrane shape, topology changes 
such as budding, equilibrium and non equilibrium fluctuations, membrane adhesion, long 
range interactions on membranes etc... It is still under development to include out of 
equilibrium features such as lipid and protein exchange with the bulk, but much has been 
learned already. 

Efforts towards understanding the cytoskeleton are more recent. They have 
been focused on the description of its passive visco-elastic properties which are now fairly 
well understood in terms of gels made of cross-linked semi-flexible polymers 
\cite{head03,wilh03}. Such 
materials which can be prepared in vitro, are equilibrium systems which obey 
conventional thermodynamics. In eukaryotic cells, the problem is however qualitatively new: 
the cross-links can be made of  motor proteins which have their own dynamics driven by 
chemical energy. Experiments, simulations and analytical descriptions, 
have shown that such systems have a rich and complex behavior \cite{taki91,nede97,surr98,nede01,surr01,seki98,krus00,krus01,krus03,krus03a,lee01,kim03,live03}. 
One can grasp the 
degree of complexity with the remark that cross-links define distances, in other words they 
define a metric. The cross-linking agents being  motor proteins they move and the metric evolves with 
time. 

In fact the problem  is 
even more complex since cytoskeletal filaments are polar and out of equilibrium: they 
polymerize at one end while depolymerizing at the other end. Such a process  called 
treadmilling is well known to biologists. The description of eukaryotic cytoskeletal gels 
should include all these features. We call these gels and more generally all gels working 
in the presence of a permanent energy consumption  "active" gels. 
On large length scales and long time scales, the properties of  complex materials
can be captured by a generalized hydrodynamic theory based on conservation
laws and symmetry considerations. Such theories have been very successful in describing complex fluids such as
liquid crystals, polymers, conventional gels, superfluids etc. \cite{mart72,dege70,seki91}.
For instance all liquid crystal display devices can be described 
by such theories. More recently active liquid crystals have been considered 
within this same logic \cite{simh02}.
Here, we develop a hydrodynamic theory
of  active gels. An example of such gel is the network of actin cytoskeletal
filaments  in the presence of chemical processes which locally
induce filament sliding and generate motion. These active processes are 
mediated by motor proteins which hydrolyze a fuel, Adenosinetriphosphate
(ATP). Since cytoskeletal filaments are structurally polar, each filament defines
a vector. This filamental structure can on large scales give rise to a
polarity of the material if filaments are aligned on average.

We develop the hydrodynamic theory of active polar gels systematically 
in several steps, following standard procedures. First, we identify the
relevant fields and write conservation laws for conserved quantities. We
identify the generalized fluxes and conjugate forces in the system. These fluxes
and forces define the rates of entropy production and dissipation.
Using the signature of forces and fluxes with respect to time reversal, we
define dissipative and reactive fluxes. These fluxes can be expressed by a
general expansion in terms of forces, by writing all terms allowed by
symmetry and by respecting the signatures with respect to time reversal.
By keeping the relevant terms to lowest order, this finally results in generic dynamic
equations which are valid in the vicinity of thermal equilibrium. We then provide
examples by analyzing the active gel mode structure and by discussing 
the spontaneous dynamical behavior of topological singularities such 
as disclinations (asters and vortices) in two dimensions. 
Eventually we discuss the merits of our approach, its current limitations and ways to extend its 
domain of validity to the more relevant far from equilibrium regime. Some of the
results discussed here have been presented in Ref. \cite{krus04}.

\section{Fields, densities and conservation laws}

We develop the  hydrodynamic description of active gels starting with conservation laws. 
The number density of monomers in the gel is denoted by $\rho$. The gel is created by 
polymerization of filaments 
from monomers. The monomers have a density $\rho^{(a)}$ in solution. The treadmilling process and the active stresses in the 
gel create a flow of the actin gel monomers with a local velocity ${\bf v}$. The gel 
current is then convective and
mass conservation implies that
\begin{equation}
\frac{\partial \rho}{\partial  t} +\nabla {(\bf v \rho)}=  -k_d \rho + k_p \delta_S\label{drhodt}
\end{equation}
Here 
$k_d$ is the depolymerization rate; we assume, as seems to be the case for actin, 
that the depolymerization which occurs mostly at the branching points of the gel, has a rate proportional
to the local density. In some cases treated below, we will consider for simplicity that depolymerization 
does not occur in the bulk but at the surface of the gel. 
In many situations, filament polymerization is highly controlled and localized
for example at the surface of the gel. This is taken into account in Eq. (\ref{drhodt})
by introducing the surface polymerization rate $k_p$ and $\delta_S$ denotes a
Dirac-like distribution which is non-vanishing only at the gel surface $S$.
Similarly, we write a conservation law for free diffusing monomers,
\begin{equation}
\frac{\partial \rho^{(a)}}{\partial  t} +\nabla {\bf j}^{(a)} =  k_d \rho -k_p \delta_S
\end{equation}
here, we have introduced the diffusive current ${\bf j}^{(a)}$ of free monomers. 

Active processes are mediated by molecular motors. The concentration of motors bound to the
gel $c^{(b)}$ is an important quantity to characterize the effects of active processes in the
gel. Assuming for simplicity that the total number of motors is conserved, we write
conservation laws for $c^{(b)}$ and the concentration of freely diffusing motors in the
solvent $c^{(m)}$ which read
\begin{eqnarray}
\frac{\partial c^{(m)}}{\partial  t} +\nabla {\bf j}^{(m)}  &=&  k_{\rm off} c^{(m)} -k_{\rm on} \rho (c^{(m)})^{n}\nonumber\\
\frac{\partial c^{(b)}}{\partial  t} +\nabla c^{(b)}{\bf v}+\nabla {\bf j}^{(b)}  &=&  - k_{\rm off} c^{(b)} +k_{\rm on} \rho (c^{(m)})^{n}\label{dcbdt}
\end{eqnarray}
The attachment and  detachment rates
of motors to and from the gel  are characterized by the chemical rates $k_{\rm on}$ and $k_{\rm off}$. 
Here, we have taken into account that bound motors are convected with the gel.
The current of
free motors  is ${\bf j}^{(m)}$ and we denote ${\bf j}^{(b)}$ the current of bound motors relative
to gel motion. 
In general, binding of motors to the gel is cooperative, and cannot be described as
a second order reaction:
groups of  motors could bind together. We use here an $n^{th}$ order chemical kinetics where the rates is proportional to  $(c^{(m)})^n$.

A final important conservation law is momentum conservation. In biological gels on scales of micrometers, inertial forces are negligible and
momentum conservation is replaced by the force balance condition which reads
\begin{equation}
\partial_\alpha (\sigma^{\rm tot}_{\alpha\beta}-\Pi\delta_{\alpha\beta})+f^{\rm ext}_\beta=0 \label{fb}
\end{equation}
where $f_\beta^{\rm ext}$ is an external force density.
Locally, there are two forces acting on the gel,
the total stress tensor $\sigma^{\rm tot}_{\alpha\beta}$ and the pressure $\Pi$.

The dynamics of the system is specified if
the flow velocity ${\bf v}$ and the currents ${\bf j}^{(a)}$, ${\bf j}^{(m)}$ and ${\bf j}^{(b)}$ are known.
The physical description of the currents is discussed in the following sections. Furthermore,
we have to take into account the polar nature of the gel. Individual filaments are rod-like objects with
two different ends which therefore have a vectorial symmetry. If the filaments in the gel 
are on average aligned, the material is oriented. 
We introduce a polarization field ${\bf p}$ to describe this orientation. The field $\bf p$ is defined
by associating with each filament a unit vector pointing to one end. The vector ${\bf p}$
is given by the local average of a large number of these unit vectors.

\section{Constitutive equations}

The fluxes of monomers and motor molecules are generated by forces
which act on the active gel and induce motion. In this section, we identify the relevant forces 
and derive general flux-force relations. These relations define the
material properties, and characterize how the system reacts to different types
of generalized forces. Of particular significance for our theory is the existence
of active processes mediated by molecular motors. In general,
a chemical fuel, such as Adenosine-triphosphate ($ATP$), is used as an energy source.
Motor molecules consume $ATP$ by catalyzing the hydrolysis to Adenosinediphosphate
($ADP$) and an inorganic phosphate and transduce the free energy of this reaction 
to generate forces and motion along the filaments. The energy of $ATP$ is also
used for polymerization and depolymerization of the filaments. The presence of the
fuel is equivalent to a chemical "force" acting on the system. We characterize this
generalized force by the chemical potential difference $\Delta\mu$ of
$ATP$ and its hydrolysis products, $ADP$ and phosphate. 
When $\Delta\mu$ vanishes the chemical reaction is at equilibrium and
there is no energy production. When $\Delta\mu >0$, a free energy 
$\Delta\mu$ is consumed per hydrolyzed $ATP$ molecule.

Constitutive equations are obtained by first identifying the fluxes and the
corresponding conjugate generalized forces and then writing 
a general expansion for the fluxes in terms of the forces. We study in this paper an active gel close to thermodynamic equilibrium and limit the 
expansion of fluxes in terms of forces to linear order as in a standard Onsager theory. We thus describe the linear response of  the 
gel to generalized forces. We do this in the most general way and write in the flux-force expansion all terms which are consistent with the symmetries
of the system. 

\subsection{Fluxes and forces}

We first discuss the rate of entropy production in the active gel.
A change in the free energy $F=U-TS$ per unit time can be written as
\begin{equation}
\dot F = -\int d^3{\bf r}\left\{\sigma^{\rm tot}_{\alpha\beta} \partial_\alpha v_{\beta} +h_\alpha  \dot p_\alpha
               +  \Delta\mu r
               + \partial_\alpha j_\alpha^{(i)} \mu^{(i)}\right\}\label{fdot}
\end{equation}
where the ``dot'' denotes a time derivative. The total deviatory stress
tensor $\sigma^{\rm tot}_{\alpha\beta}$
is in general is not symmetric; it is conjugate to the velocity gradient
$\partial_\alpha v_\beta$. The field conjugate to  the order parameter $p_\alpha$ is the functional derivative of the free energy $F$ of the 
gel at thermal equilibrium,
$h_\alpha=-\frac{\delta F}{\delta p_\alpha} $, where the functional derivative is taken for constant
deformation, temperature and number of particles. 
The current conjugate to the field $h_\alpha$ is the convected time derivative of the polarization $\dot p_\alpha =\frac{\partial p_\alpha}{\partial t} +v_\beta \partial_\beta p_\alpha$. The chemical force
$\Delta \mu$ is conjugate to the $ATP$ consumption rate $r$ which determines the
number of $ATP $ molecules hydrolyzed per unit time and per unit volume.
Finally, $\mu^{(a)}$, $\mu^{(m)}$ and $\mu^{(b)}$ are the chemical potentials
of free monomers, free motors and of motors bound to the gel, respectively.

Eq. (\ref{fdot}) does not take into account the translational and rotational invariance of the active gel,
and the variables used are therefore not the proper conjugate fluxes
and forces. Indeed, as shown in reference \cite{dege93}, since the free energy does not change under pure translations
and rotations of the gel, ignoring surface terms, we can rewrite Eq. (\ref{fdot}) as
\begin{equation}
\dot F = -\int d^3{\bf r}\left\{\sigma_{\alpha\beta} u_{\alpha\beta} +h_\alpha  P_\alpha 
+  \Delta\mu r
               - j_\alpha^{(i)}\partial_\alpha  \mu^{(i)}\right\}\label{fdot1}
\end{equation}
The total stress $\sigma^{\rm tot}$ has been decomposed into a symmetric part $\sigma_{\alpha\beta}$ (with $\sigma_{\alpha\beta}=\sigma_{\beta\alpha}$) and and anti-symmetric part which is due to the torque exerted by the field $h_\alpha$ on the order parameter $p_\alpha$:
\begin{equation}
\sigma^{\rm tot}_{\alpha\beta}=\sigma_{\alpha\beta}+\frac{1}{2}(p_\alpha h_\beta - p_\beta h_\alpha)
\end{equation}
Here, $u_{\alpha\beta} = \frac{1}{2}(\partial_\alpha v_\beta + \partial_\beta v_{\alpha})$
is the symmetric part of the  velocity gradient tensor.
The current conjugate to $h_\alpha$  includes a rotational contribution coming from the antisymmetric part of the stress tensor, it reads
\begin{equation}
P_\alpha=\frac{D}{Dt}p_\alpha=\frac{\partial p_\alpha}{\partial t}+(v_\gamma\partial_\gamma)p_\alpha
+\omega_{\alpha\beta} p_\beta
\end{equation}
where we have used the corotational time derivative of the vector $p_\alpha$,
$\omega_{\alpha\beta}=\frac{1}{2}(\partial_\alpha v_\beta - \partial_\beta v_{\alpha})$
being  the vorticity tensor of the flow. 

We can read off Eq. (\ref{fdot1}) the following pairs of
conjugate fluxes and forces:
\begin{eqnarray}
\textrm{flux}&\leftrightarrow&\textrm{force}\nonumber \\
 \sigma_{\alpha\beta} &\leftrightarrow & u_{\alpha\beta}\nonumber\\
P_\alpha &\leftrightarrow & h_\alpha\\
r&\leftrightarrow &\Delta\mu\nonumber\\
j_\alpha^{(i)}&\leftrightarrow & \partial_\alpha \mu^{(i)}\nonumber
\end{eqnarray}

The rate of change of the free energy can be divided into reversible and irreversible
parts $\dot F=\dot F_{\rm rev}+\dot F_{\rm irr}$ where $\dot F_{\rm rev}=\dot U$
and $\dot F_{\rm irr}=-T \dot S$.
We therefore
decompose the fluxes into a reactive part and a dissipative part. They are characterized by
their different signatures with respect to time-reversal. Note, that the generalized forces
have well-defined signatures with respect to time reversal: the rate of strain 
$u_{\alpha \beta}$ is odd under time reversal, while $h_\alpha$ and $\Delta\mu$
are even. We write
\begin{eqnarray}
\sigma_{\alpha\beta} & = & \sigma_{\alpha\beta}^r+\sigma_{\alpha\beta}^d \nonumber\\
P_\alpha & = & P_\alpha^r+P_\alpha^d \nonumber\\
r  & = &r^r+r^d\quad.
\end{eqnarray}
The dissipative fluxes
have the same signature under time reversal as their conjugate forces, while
reactive fluxes have the opposite signature under time reversal.
Therefore, $\sigma^d_{\alpha\beta}$ is odd under time reversal, while
$r^d$ and $P_\alpha^d$ are even. The reactive parts have correspondingly 
the opposite signatures. As we show below, the currents $j^{(i)}_\alpha$ do not have reactive parts.
With this decomposition,  the rate of  entropy production $T\dot S=-\dot F_{\rm irr}$
is given by
\begin{equation}
\label{eq:diss}
T\dot{S} = \int d^3{\bf r}\left\{r^d\Delta\mu+P_\alpha^d h_\alpha 
               + \sigma^d_{\alpha\beta}u_{\alpha\beta} - j_\alpha^{(i)}\partial_\alpha\mu^{(i)}\right\}\quad.
\end{equation}
For a changing state of the system which is periodic in time with period $T$, 
the internal energy is the same after one
period. Therefore, in this case
\begin{equation}
\int_0^T dt(r^r\Delta\mu + P_\alpha^r h_\alpha + \sigma^r_{\alpha\beta}u_{\alpha\beta})=0\quad.
\end{equation}
These relations follow from the signatures of dissipative and reactive currents
under time reversal.

\subsection{Maxwell model}

In the absence of a net gel polarization ${\bf p}$ and for a passive system with
$\Delta\mu=0$, we describe the viscoelastic gel by a Maxwell model. In order to keep
our expressions simple and to focus on the essential physics, we assume that the
ratio of the bulk and shear viscosities is $2/d$ where $d$ is the space dimension. 
The Maxwell model relates the stress tensor
to the strain rate according to
\begin{equation}
\left (1+\tau \frac{D}{Dt}\right )\sigma_{\alpha\beta}=2\eta u_{\alpha\beta}\label{maxw}
\end{equation}
Here, $\eta$ denotes the shear viscosity and $\tau$ the viscoelastic relaxation time.
and $E=\eta/\tau$ is the elastic modulus of the gel at short times.
In Eq. (\ref{maxw}), the time derivative of the stress tensor taken in a reference frame
moving
with the material flow must be used.  
For a tensor, this implies that in the laboratory frame convective terms due to the fluid flow
as well as the rotation
of the reference system due to the vorticity of the flow 
need to be taken into account.
 $D/Dt$ denotes the corotational derivative of a tensor given by
\begin{equation}
\frac{D \sigma_{\alpha\beta}}{Dt}\equiv \frac{\partial\sigma_{\alpha\beta}}{\partial t} +
(v_\gamma \partial_\gamma)\sigma_{\alpha\beta}
+\left[\omega_{\alpha\gamma}\sigma_{\gamma\beta}
+\omega_{\beta\gamma}\sigma_{\gamma\alpha}\right]\label{eq:convectedDerivative}
\end{equation}
Note that we include the geometrical non-linearities and that we use the most general versions of what is called a ``Convected Maxwell Model''
\cite{bird87}.
For the Maxwell model given by Eq. (\ref{maxw}), we can specify the reactive and the
dissipative contributions to the stress tensor.
\begin{eqnarray}
(1-\tau^2 \frac{D^2}{Dt^2})\sigma_{\alpha\beta}^d&=&2\eta u_{\alpha\beta}\label{sd} \\
\sigma_{\alpha\beta}^r&=&-\tau \frac{D}{Dt}\sigma_{\alpha\beta}^d\label{sr}
\end{eqnarray}
With these relations, we have $\sigma_{\alpha\beta}=\sigma_{\alpha\beta}^d+\sigma_{\alpha\beta}^r$
and the dissipative and reactive parts of the stress differ in their signature with respect to
time reversal. 

\subsection{Dissipative fluxes}

We now generalize the relations (\ref{sd}) and (\ref{sr}) expressing dissipative
and reactive fluxes in terms of generalized forces for an active polar gel. First,
we write expressions for  the dissipative fluxes. To linear order we find
\begin{eqnarray}
\label{eq:stressDissipative}
\left(1-\tau^2\frac{D^2}{Dt^2}\right) \sigma_{\alpha\beta}^d  &=&  2\eta u_{\alpha\beta}\nonumber \\
  &-& \tau \frac{D}{Dt}(\frac{\nu_1}{2} ( p_\alpha h_\beta+p_\beta h_\alpha)+\bar\nu_1 p_\gamma h_{\gamma}  \label{sigmadab}\\
\left(1-\tau^2\frac{D^2}{Dt^2}\right) P^d_\alpha& =& \left(1-\tau^2\frac{D^2}{Dt^2}\right) ( \frac{h_\alpha}{\gamma_1} + \lambda_1p_\alpha \Delta\mu)\nonumber \\
 &+&\tau \frac{D}{Dt}(\nu_1 p_\beta u_{\alpha\beta}+\bar\nu_1 p_\alpha u_{\beta\beta})\label{pda}\\
r^d  &=&  \Lambda\Delta\mu + \lambda_1p_\alpha h_\alpha + \lambda p_\alpha \partial_\alpha \mu^{(b)} \quad.
\end{eqnarray}
Note that the first expression is a generalization 
for  Eq. (\ref{sd}) of the Maxwell model.  For a sake of simplicity, we ignore the anisotropy of the viscosity and 
consider that all translational viscosities are equal. This anisotropy could be introduced as in the classical 
description of the hydrodynamics of liquid crystals. Couplings to $\Delta\mu$ cannot appear in the equation 
of the dissipative stress 
because it transforms differently under time reversal than $u_{\alpha\beta}$.
A term coupling $\sigma_{\alpha\beta}^d$ 
to the time derivative of $h_\alpha$ can appear. It  is required by 
Onsager symmetry relations since a corresponding 
coupling term with coefficient $\nu_1$ does occur in the equation for $P^d_\alpha$
as shown in Appendix B.
The dissipative coefficients
$\gamma_1$ and $\Lambda$ characterize the coupling of $P^d_\alpha$ and $r^d$ to
$h_\alpha$ and $\Delta\mu$, respectively. 
Here,  $\gamma_1$ is the rotational viscosity which appears in classical 
nemato-hydrodynamics. Because of its signature with respect to time reversal,
cross-coupling terms involving $u_{\alpha\beta}$ do not appear
in the expression $r^d$. The expression for $P^d_\alpha$ contains a terms coupling $P^d_\alpha$ 
to the time derivative of $u_{\alpha\beta}$  as derived in Appendix B. Cross-coupling terms occur also
in the expressions for $P^d_\alpha$ and $r^d$. Since $P_\alpha$ is a vector and $r$ a scalar,
we need a vector in the system to couple these equations. If the system is polar, this  vector
is provided by ${\bf p}$. Therefore, the cross-coupling term characterized by $\lambda_1$
involves $p_\alpha$. Note that
because of the symmetry of the Onsager coefficients, the coefficients $\lambda_1$
$\nu_1$ and $\bar \nu_1$
are the same in different equations.
The last term in the expression of $r^d$ is a crossed term with the current of bound motors discussed below. 
This term is also imposed by the Onsager symmetry relations.
We still have to specify the currents $j^{(a)}_\alpha$, $j^{(m)}_\alpha$ and $j^{(b)}_\alpha$
of monomers and motor molecules
which do not have reactive components:
\begin{eqnarray}
j_{\alpha}^{(a)}&=&-D^{(a)}\partial_\alpha \rho^{(a)}\\
j_\alpha^{(m)}&=&-D^{(m)}\partial_\alpha c^{(m)}\\
j_\alpha^{(b)}  & = & -D^{(b)} \partial_\alpha c^{(b)} +\lambda p_\alpha\Delta\mu\label{jb}
\end{eqnarray}
The first two expressions are simple diffusive currents with diffusion coefficients $D^{(i)}$, where
we have assumed for simplicity that couplings
to $h_\alpha$ can be neglected and that free diffusion is not influenced by $\Delta\mu$.
The expression (\ref{jb}) describes bound motors which are convected with the gel. In addition to being 
convected with velocity ${\bf v}$, motors move in a direction given by
the filament orientation if $\Delta\mu$ is nonzero. This directed motion is characterized
by the coefficient $\lambda$; the motor  velocity $v^m$ on the filaments is such that
$c^{(b)}  v^m_{\alpha}=\lambda p_\alpha\Delta\mu$.
 Finally, velocity fluctuations of motor motion are captured
by a diffusive term with diffusion constant $D^{(b)}$.

\subsection{Reactive fluxes}

Reactive fluxes are expressed in terms of the forces following their signature
under time translation invariance. Generalizing Eq.(\ref{sr}) for the reactive stresses
of the Maxwell model, we note that here cross coupling terms with the forces $h_\alpha$ and
$\Delta\mu$ are possible. Taking into account the tensorial structure of the stress tensor, we
write 
\begin{eqnarray}
\sigma_{\alpha\beta}^r & = & -\tau\left[\frac{D\sigma_{\alpha\beta}^d}{Dt} + A_{\alpha\beta}
\right]
 -\zeta\Delta\mu p_\alpha p_\beta-\bar\zeta\Delta\mu\delta_{\alpha\beta}\nonumber\\
&-&\zeta'\Delta\mu p_\gamma p_\gamma\delta_{\alpha\beta}\nonumber\\
&+&\frac{\nu_1}{2}(p_\alpha h_\beta+p_\beta h_\alpha) + \bar\nu_1 p_\gamma h_\gamma\delta_{\alpha\beta}\quad,\label{sigmar}
\end{eqnarray}
where we have introduced the phenomenological coefficients $\zeta$, $\bar\zeta$, $\zeta'$,
$\nu_1$
and $\bar\nu_1$. The tensor
\begin{eqnarray}
A_{\alpha\beta} &=& \nu_2(u_{\alpha\gamma}\sigma_{\gamma\beta}^d+
\sigma_{\alpha\gamma}^du_{\gamma\beta})+ 
\nu_3u_{\gamma\gamma}\sigma_{\alpha\beta}^d + \nu_4u_{\gamma\gamma}\sigma_{\delta\delta}^d\delta_{\alpha\beta} \nonumber \\
&+& \nu_5\sigma_{\gamma\gamma}^du_{\alpha\beta} + \nu_6 u_{\gamma\delta}\sigma_{\delta\gamma}^d\delta_{\alpha\beta} \quad.
\label{eq:A}
\end{eqnarray}
contains nonlinear reactive terms to lowest order, resulting from the geometry of the flow field
with corresponding phenomenological coefficients $\nu_i$. 
Similar 
coefficient have been introduced in the so-called ``eight constant Oldroyd model'' in rheology 
\cite{bird87}.
The term proportional to $\tau$ on the right hand side of Eq. (\ref{sigmar})
assures compatibility with the Maxwell model in the absence of polarization and
chemical fuel. 
For the reactive parts of $P_\alpha$ and $r$ we write
\begin{eqnarray}
(1-\tau^2\frac{D^2}{Dt^2}) P_\alpha^r & = &  - \nu_1u_{\alpha\beta}p_\beta-\bar\nu_1u_{\beta\beta}p_\alpha\label{Pr} \\
r^r&=&\zeta p_\alpha p_\beta u_{\alpha\beta}+\bar\zeta u_{\alpha\alpha}+ \zeta' p_\alpha p_\alpha u_{\beta\beta}
\label{rr}
\end{eqnarray}
Here we have written cross coupling terms of $P^r$ with the rate of strain
$u_{\alpha\beta}$. The linear response matrix of reactive terms is antisymmetric, therefore the same
coefficients as in Eq. (\ref{sigmar}) appear, however with opposite sign.
The remaining Eq. (\ref{rr}) is constructed in the same way with cross coupling coefficients
that have been introduced in Eq. (\ref{sigmar}). No reactive cross-terms between
Eqns. (\ref{Pr}) and (\ref{rr}) exist. 
 
\subsection{Dynamic equations}

Using the expressions for dissipative and reactive fluxed discussed above, we
now write general hydrodynamic equations for the active viscoelastic and
polar gel. Adding the dissipative and reactive parts, we find
\begin{eqnarray}
2\eta u_{\alpha\beta} & = & 
\left(1+\tau\frac{D}{Dt}\right)\left\{\sigma_{\alpha\beta} 
+ \zeta\Delta\mu p_\alpha p_\beta \right .\nonumber \\
& & \left . + \zeta'\Delta\mu p_\gamma p_\gamma\delta_{\alpha\beta}+
\bar\zeta\Delta\mu\delta_{\alpha\beta}+ 
\tau A_{\alpha\beta} \right\}
 \nonumber\\
&-&   \frac{\nu_1}{2}(p_\alpha h_\beta+p_\beta h_\alpha) - 
\bar\nu_1 p_\gamma h_\gamma\delta_{\alpha\beta}  
\label{uab}\\
\left(1-\tau^2\frac{D^2}{Dt^2}\right)\frac{D p_\alpha}{D t}  &=& \left(1-\tau^2\frac{D^2}{Dt^2}\right)( \frac{1}{\gamma_1} h_\alpha 
+ \lambda_1p_\alpha \Delta\mu )\nonumber\\
& & - (1-\tau \frac{D}{Dt})(\nu_1u_{\alpha\beta}p_\beta+
\bar\nu_1u_{\beta\beta}p_\alpha) \label{eq:dpdt}\\ 
r &=& \zeta p_\alpha p_\beta u_{\alpha\beta}+\bar\zeta u_{\alpha\alpha} + \zeta' p_\alpha p_\alpha u_{\beta\beta}\nonumber\\
 &+& \Lambda\Delta\mu 
+ \lambda_1p_\alpha h_\alpha\quad. \label{r}
\end{eqnarray}
In these equations, we have included geometric nonlinearities but have restricted
other terms to linear order for simplicity. Also, we have neglected chiral
terms which in principle exist in cytoskeletal systems but which are expected to
be small. These equations are complemented by the force balance condition
(\ref{fb}).

Eq. (\ref{uab}) generalizes the expression of the stress tensor in the 
Maxwell model to active systems with polarity. Indeed, even in the absence of
stresses, the active terms proportional to $\Delta\mu$ generate a finite strain rate.
Similarly, if all flows are suppressed, the active terms generate a nonzero stress
tensor. Thus, the hydrolysis of $ATP$ can generate forces and material flow in the
gel via the action of active elements such as motors. These terms are characterized by the
coefficients $\zeta$, $\zeta'$ and $\bar \zeta$ 
Similarly, we find
active terms in the polarization dynamics given by Eq. (\ref{eq:dpdt}) described by the coefficient
$\lambda_1$. Furthermore, material flow couples to the polarization dynamics via
the coefficients $\nu_1$ and $\bar\nu_1$. The rate of $ATP$ consumption $r$ is primarily  driven
by $\Delta\mu$ and characterized by $\Lambda$. However, it is also coupled to the fluid
flow and to the field ${\bf h}$ acting on $\bf p$.
Note, that in addition surface terms can be important. For example, if filaments polymerize
at the gel surface with a rate $k_p$ (see Eq. (\ref{drhodt})), there is an additional contribution
$r_S$ to ATP consumption  which is localized at the surface. In the following, we discuss situation 
where surface effects can be captured by introducing appropriate boundary conditions.

\section{Hydrodynamic modes}

We now calculate the linear relaxation modes of an active polar gel. 
They can be obtained from the linear response of the change in the gel density to
externally applied forces.
We focus on the situation of a fully polarized gel far from an isotropic polar transition
with the polarization
vector lying in the $x$-direction, i.e., ${\bf p}={\bf e}_x$ is constant. 
We assume that the gel is treadmilling, which implies that it moves at
constant velocity ${\bf v}=v_0{\bf e}_x$, while the gel mass is conserved. The
treadmilling is due to polymerization and depolymerization
of the filaments at the gel surfaces located at $x=x^+$ and $x=x^-$, respectively.
In the following, we consider the case of an infinite system where $x^+$ are located at
$+\infty$ and $x^-$ to $-\infty$ while the treadmilling velocity  
$v_0$ remains constant.

We restrict ourselves to
modes for which the fields vary along 
the $x$-direction which is the treadmilling direction, while the system is assumed to be 
homogeneous in the other directions. In this case the dynamic equations 
(\ref{drhodt})-(\ref{dcbdt}) for the densities of polymerized actin, bound and free myosin become
effectively one-dimensional and read
\begin{eqnarray}
\partial_t\rho+\partial_x(\rho v) & = & 0\\
\partial_t c^{(m)} -D\partial_x^2 c^{(m)}  &=&  k_{\rm off} c^{(b)} -k_{\rm on}
\rho (c^{(m)})^{n}\label{eq:cm}\\
\partial_tc^{(b)} + \partial_x(c^{(b)}v+\lambda \Delta\mu) & = & - k_{\rm off} c^{(b)} +k_{\rm on} \rho (c^{(m)})^{n}\quad.
\end{eqnarray}
The actin monomers which are not part of the gel diffuse across the
system and control by their concentrations the treadmilling rate.
The mechanical constitutive equation (\ref{uab}) takes the
form 
\begin{equation}
\label{eq:s}
2\eta\partial_xv =  [1+\tau\partial_t+\tau v\partial_x ](\sigma+\zeta\Delta\mu) 
\end{equation}
whereas the force balance in the presence of an external stress 
$\sigma^{\rm ext}$ is 
expressed by
\begin{equation}
\label{eq:f}
\partial_x(\sigma+\sigma^{\rm ext})  =  \partial_x\Pi\quad.
\end{equation}
In these equations, the coefficients $\lambda$
and $\zeta$, which  are the Onsager coefficients introduced
in the previous section, characterize the 
directed motion of motors along the filaments and the actively generated stress in the gel
(we have replaced here  $\zeta+\zeta'+\bar\zeta$ by $\zeta$.).
In a situation far from equilibrium, nonlinearities
become important. We therefore linearize around the homogeneous
steady state of the above equations given by
$\rho=\rho_0$, $c^{(b)}=c^{(b)}_0$, 
$c^{(m)}=c^{(m)}_0$,  $v=v_0$, and $\sigma=\sigma_0$, where 
$k_{\rm off} c^{(b)}_0 =k_{\rm on} \rho_0 (c^{(m)}_0)^{n}=\rho_0 v_0$
 and $\sigma_0=-\zeta\Delta\mu$. We consider 
a small perturbation of this state, i.e., 
$\rho=\rho_0+\delta\rho$ and analogously for the remaining quantities.
To leading order, the active coefficients can be expanded as
$\zeta=\zeta_0+\zeta_\rho\delta \rho+\zeta_c \delta c^{(b)}$ and $\lambda=\lambda_0+
\lambda_\rho \delta \rho 
+\lambda_c \delta c^{(b)}$ to implement such nonlinearities which become relevant
in all realistic situations. Here, we have neglected the diffusion of bound motors and the 
diffusion constant $D^{(b)}$ has been set
to zero.

The linearized dynamical equations can be solved by
Fourier-transformation in space and time. The mode characterizing the
variations of the density of unbound  motors, $c^{(m)}(q,\omega)$ as a function of spatial
wave vector $q$ and temporal frequency $\omega$, can be obtained from 
Eq.~(\ref{eq:cm}) and is given by
\begin{equation}
c^{(m)}(q,\omega) = \frac{k_{\rm off}c^{(b)}(q,\omega)-k_{\rm on}(c^{(m)}_0)^n\rho(q,\omega)}
{i\omega+Dq^2+T_c^{-1}}\quad,
\end{equation}
where $T_c^{-1}= nk_{\rm on}\rho_0(c^{(m)}_0)^{n-1}$ is a chemical relaxation time.
Using this result as well as the linearized conservation equation 
\begin{equation}
i\omega\rho(q,\omega) + iq\rho_0v(q,\omega)+iqv_0\rho(q,\omega)=0
\end{equation}
we obtain for the distribution of bound motors
\begin{equation}
c^{(b)}(q,\omega) = \frac{A(q,\omega)c_0-
iq\bar\lambda_\rho}{A(q,\omega)+iq\bar\lambda_c}
\frac{\rho(q,\omega)}{\rho_0}
\end{equation}
where $\bar\lambda_\rho = \lambda_\rho \Delta\mu\rho_0$,
$\bar\lambda_c=\lambda_c \Delta\mu$, and
\begin{equation}
A(q,\omega) = i\omega + iqv_0+\frac{i\omega+Dq^2}{i\omega+Dq^2+T_c^{-1}}k_{\rm off}\quad.
\end{equation}
The mechanical equations (\ref{eq:s}) and (\ref{eq:f}) then lead to
\begin{equation}
\sigma^{\rm ext}(q,\omega) = 2\mu(q,\omega)\frac{\rho(q,\omega)}{\rho_0}
\end{equation}
where the modulus $\mu(q,\omega)$ characterizes the effective material properties
in response to density variations. It  can be determined from the Eq.
\begin{eqnarray}
&&(1+i\tau(\omega+qv_0))\left[2\mu(q,\omega)-\chi\rho_0\right] = 
2iE\tau(\omega+qv_0)\nonumber\\
&&+(1+i\omega\tau+iq v_0\tau)\left[\bar\zeta_\rho+\bar\zeta_c \frac{A(q,\omega)c_0-iq\bar\lambda_\rho}{A(q,\omega)+iq\bar\lambda_c}\right]\quad.
\end{eqnarray}
Here, we have used the equation of state and have defined the inverse compressibility  $\chi = \frac {\partial \Pi}{\partial \rho}$. Furthermore, we have introduced
$\bar\zeta_\rho=\zeta_\rho\Delta\mu\rho_0$ as well as $\bar\zeta_c=\zeta_c\Delta\mu$.

The relaxation modes of the system can be identified by the (complex) values of the
frequency $\omega_n$, for which  the modulus vanishes
$\mu(q,\omega_n)=0$. Indeed, the density relaxes as 
$\rho(q)\sim e^{i\omega_n t}$ and instabilities in the system occur if the real part
of $i\omega_n$ becomes positive. For our present system, we find that three independent
relaxation modes exist, with $n=1$ to $3$.
Up to leading order in $q$, the corresponding dispersion relations are given by
\begin{eqnarray}
i\omega_1 & = & icq+dq^2\\
i\omega_2 & = & -T_c^{-1}-k_{\rm off}\\
i\omega_3 & = & -\frac{1}{\tau}
\frac{\chi\rho_0+\bar\zeta_c c_0+\bar\zeta_\rho}{2E+\chi\rho_0+\bar\zeta_c c_0+\bar\zeta_\rho}
\end{eqnarray}
The first mode is
propagating
with 
velocity
 $c$ given by
\begin{equation}
c  =  -\frac{1}{1+T_ck_{\rm off}}\left[v_0+\frac{(\chi\rho_0+\bar\zeta_\rho) 
\bar\lambda_c-\bar\zeta_c\bar\lambda_\rho}{\chi\rho_0+\bar\zeta_\rho+\bar\zeta_c c_0}\right]
\end{equation}
Note, that the propagation velocity $c$ is proportional to the fraction of 
bound motors $(1+T_ck_{\rm off})^{-1}$. The value of the coefficient $d$ can be either positive 
or negative, depending on the value of the parameters.
If it is positive, the gel is unstable and self-organizes into
propagating density profiles. Such solitary waves have been found in theoretical
descriptions of active polar fibers \cite{krus01,krus03}. Experimentally, 
actin waves have been recently reported
in several cell  types \cite{vick02,bret04,gian04}. 
 In the present context, there 
is only one propagating mode, because space
inversion symmetry has been 
broken due to treadmilling and to the polarization of the gel. The polarization couples to the filament and
motor densities
only through the active current of bound motors. If instead of motors, we have passive cross-linkers, i.e., 
$\lambda=0$ 
and in the absence of treadmilling, i.e., $v_0=0$, space inversion symmetry is restored. In this case
$c=0$ and 
\begin{equation}
d=-D\frac{k_{\rm off}}{T_c^{-1}+k_{\rm off}}\quad,
\end{equation}
so that the mode is no longer propagating, but diffusive and always stable. Note, that 
$d$ is proportional
to the fraction of unbound motors $T_ck_{\rm off} /(1+T_ck_{\rm off})$.
The existence of propagating 
waves even in a viscous environment seems to be a general feature of
active media and has already been reported in an other context by 
Ramaswamy and coworkers \cite{rama00,simh02}.

The second mode is a chemical relaxation mode describing the binding and unbinding of molecular motors 
to the polymerized gel, whereas the third mode describes the stress relaxation 
towards its stationary value. 

\section{Dynamic point defect in two dimensions}

As an example of active behavior in two dimensions described by  Eqns. (\ref{r}),
we consider in this section the dynamics of point defects in the vector
field ${\bf p}$. First, we consider the passive equilibrium state with
$\Delta \mu=0$ where all fluxes vanish.
Subsequently, we determine stationary active solutions for finite $\Delta\mu$
and determine their stability. We obtain a complete
diagram of states for asters, vortices and rotating spirals.

\subsection{Asters and vortices at thermodynamic equilibrium}

In order to study point defects in two dimension, we consider for
simplicity the situation where the orientation of ${\bf p}$
varies but the modulus is constant and we impose 
${\bf p}^2=1$. In this case, 
the free energy is given by the standard expression for a polar liquid crystal \cite{dege93}:
\begin{eqnarray}
F&=&\int d^2x \left[ \frac{K}{2}(\nabla\cdot {\bf  p} )^2 +\frac{K+\delta 
K}{2}({\bf p}\cdot\nabla{\bf p} )^2 
\right. \nonumber\\
 &+& \left. k \nabla\cdot {\bf p}-\frac{1}{2}h_\parallel {\bf p}^2 \right]
\end{eqnarray}
where $K_1=K$ and $K_3=K+\delta K$ are the splay and bend
elastic moduli and we have introduced a Lagrange multiplier $h_{||}$
to impose the constraint ${\bf p}^2=1$. 
The coefficient $k$ describes the spontaneous splay
allowed by symmetry for  vector order. Note that there is no twist term
in two dimensions.

For $\Delta\mu=0$, the system reaches an equilibrium steady state
where all fluxes and forces vanish: $\sigma_{\alpha\beta}=0$, $u_{\alpha\beta}=0$,
$P_\alpha=0$, $h_\alpha=0$ and $r=0$. The equilibrium orientation of the polarization
${\bf p}$, corresponds to a vanishing orientational field $h_\alpha=-\delta F/\delta p_\alpha=0$.
In order to describe point defects, it is convenient to introduce polar
coordinates $(r,\theta)$ and the angle $\psi(r,\theta)$ 
which characterizes
the components of the vector ${\bf p}$: $(p_r  = \cos\psi$, $p_\theta = \sin\psi)$, see Fig. 1(a).
The components of the orientational field $h_\alpha$ in cylindrical coordinates
can be written as $h_r = h_\|\cos\psi -h_\perp\sin\psi$, and $h_\theta = h_\|\sin\psi + h_\perp\cos\psi$,
where we have introduced the components $h_{||}$ and $h_\perp$
parallel and perpendicular to the direction of ${\bf p}$.

Considering for simplicity rotationally symmetric fields describing
point defects with $\psi=\psi(r)$, we obtain
\begin{eqnarray}
F &=& 2\pi\int
dr\;r\left\{\frac{K}{2}\left(\frac{1}{r}\frac{d}{dr}r\cos\psi\right)^2 \right.\nonumber\\
 &+ & \left.\frac{K+\delta K}{2}\left(\frac{1}{r}\frac{d}{dr}r\sin\psi\right)^2\right\}\quad.
\end{eqnarray}
where we have ignored the spontaneous splay which leads to a boundary
term. 

The
perpendicular field is obtained by taking the functional derivative of the free energy $h_\perp=-\delta F/\delta \psi$
\begin{eqnarray}
h_\perp&=&(K+\delta K\cos^2\psi)\left[ \psi''+\frac{\psi'}{r} 
\right]\nonumber \\
&-&\frac{\delta K}{2}\sin2\psi\left[ \frac{1}{r^2}+\psi'^2 
\right]\label{hperp}
\end{eqnarray}
where the prime indicates derivation with respect to $r$. 
The value of the Lagrange multiplier $h_\|$ (the longitudinal field) is chosen in such a way 
that  the condition ${\bf p}^2=1$ is satisfied.

Four types of topological defects of charge one are possible, see Fig. 1(a) and (b).
If we assume that boundary
conditions are always chosen to allow for solutions with constant $\psi$,
these solutions corresponds to asters with $\psi(r)=0$ and $\psi(r)=\pi$
and to vortices with $\psi(r)=\pm \pi/2$. 
The linear stability of asters and
vortices against angular perturbations $\delta \psi(r)$
is described by
\begin{eqnarray} \label{eq:fleche}
\frac{\gamma_1}
{1+\gamma_1(\nu_1\pm1)^2/(4\eta)} \frac{\partial \delta \psi}{\partial t}& =& \nonumber\\
(K +\delta K  \cos^2 \psi&)&
\left(\frac{d^2}{dr^2}+\frac{1}{r}\frac{d}{dr}\right)\delta\psi\nonumber\\
-& \delta K& \cos(2\psi)\frac{1}{r^2}
\delta \psi(r)
\end{eqnarray}
where the minus sign corresponds to an aster and teh plus sign to a vortex.
If all the eigenvalues of the linear operator on the right hand side are negative, the defect is
stable, and otherwise it is unstable.
(We may note that this operator has the form of the negative
a quantum Hamiltonian of one particle.)
Asters are stable for positive 
$\delta K$ while vortices are stable for $\delta K$ negative. For the
special case $\delta K=0$, 
Eq. (\ref{eq:fleche}) becomes $\gamma_1 \partial\psi/\partial t =K \Delta \psi$,
where $\Delta$ denotes the Laplace operator,
and defects for any value $\psi(r)=\psi_0$
are stable. These defects are called spirals
since the direction of polarization follows spirals with an equation in polar coordinates given by 
\begin{equation}
r(\theta)=r_0 \exp [ \cot (\psi_0)\theta ]\quad ,
\end{equation}
see Fig 1. (c).

\subsection{Non-equilibrium steady states}

We now discuss the effect of a point defect in two dimensions if 
motors in the system are active and $\Delta\mu\neq0$. In this situation,
spiral defects start to rotate and there are nontrivial flow and stress profiles. 
We assume that the system is incompressible, i.e.,
$u_{rr}+u_{\theta\theta}=0$. Using the expression of the  strain rate tensor given in appendix, this imposes that 
$v_r=\alpha/r$. The absence of singularity of the radial velocity for small $r$ implies that
$\alpha=0$, therefore $v_r=0$ and $u_{rr}=u_{\theta\theta}=0$.

The Onsager Eq. (\ref{eq:dpdt}) for the polarization rate can then be written in cylindrical coordinates. 
In a steady state, $P_\alpha=0$, we obtain at linear order:
\begin{eqnarray}
\nu_1u_{r\theta}\sin2\psi & = & \frac{h_\|}{\gamma_1} +
\lambda_1\Delta\mu\label{eq:hParallel}\\
(\nu_1\cos2\psi - 1)u_{r\theta} & = &
\frac{h_\perp}{\gamma_1}\quad.\label{eq:hPerp}
\end{eqnarray}
 where we have expressed the orientational field $h_\alpha$ in terms of its parallel and perpendicular coordinates.
Similarly,  Eqns. (\ref{uab})-(\ref{r}) can be rewritten in polar coordinates.
Using $u_{rr}=u_{\theta\theta}=0$ and only taking terms to linear
order in the generalized forces into account we find that the
steady state obeys
\begin{eqnarray}
\label{eq:uRR}
0 & = & \sigma_{rr}
+\zeta\Delta\mu \cos^2\psi +
(\bar\zeta+\zeta')\Delta\mu\nonumber\\
& & -\nu_1\left[h_\|\cos^2\psi-\frac{1}{2}h_\perp\sin2\psi \right]
\nonumber\\
& &-\bar\nu_1h_\| \\
\label{eq:uTT}
0 & = & \sigma_{\theta\theta} +
\zeta\Delta\mu\sin^2\psi+
(\bar\zeta+\zeta')\Delta\mu\nonumber\\
& & -\nu_1\left[h_\|\sin^2\psi+\frac{1}{2}h_\perp\sin2\psi \right]
\nonumber\\
& &-\bar\nu_1 h_\| \\
\label{eq:uRT}
2\eta u_{r\theta} & = & \sigma_{r\theta} +
\frac{\zeta}{2}\Delta\mu\sin2\psi \nonumber\\
& &
-\frac{1}{2}\nu_1\left[h_\|\sin2\psi+h_\perp\cos2\psi\right]\quad.
\end{eqnarray}
In addition to the dynamic equations Eqns. (\ref{uab})-(\ref{r}), 
the force balance Eq. (\ref{fb}) has to be satisfied.
The $\theta$ component of the force balance equation 
( Eq. (\ref{srt}) of appendix \ref{sec:app1}) is solved by
$\sigma^{\rm tot}_{r\theta}=B/r^2$ where $B$ is an integration constant. 
Since $\sigma^{\rm tot}_{r\theta}$ must not diverge, $B=0$ and therefore $\sigma^{\rm tot}_{r\theta} = 0$
and $\sigma_{r\theta}=h_\perp/2$.
The components $\sigma_{rr}$ and $\sigma_{\theta\theta}$ of the
stress tensor follow from Eqns. (\ref{eq:uRR}) and (\ref{eq:uTT}).
These stresses together with the radial component of the force balance equation (Eq. (\ref{srrtt}) of appendix \ref{sec:app1}) determine the pressure
profile $\Pi$ required to ensure the incompressibility of the material.

\subsection{Asters, vortices and spirals}

We now determine the stationary solutions for point defects in non-equilibrium
states. From Eq.~(\ref{eq:hParallel}) we find an expression for the
parallel component of the orientational field
\begin{equation}
h_\| = \gamma_1(\nu_1 u_{r\theta}\sin2\psi-\lambda_1\Delta\mu)\quad.
\end{equation}
Inserting this expression in Eq.~(\ref{eq:uRT}), we obtain the non diagonal strain rate
\begin{equation}
u_{r\theta}=\frac{\tilde\zeta\Delta\mu\sin 2\psi 
+ h_\perp(1-\nu_1\cos 2\psi)}{4\eta + \gamma_1\nu_1^2\sin^2 2\psi}\label{urt0}
\end{equation}
where $\tilde\zeta=\zeta+\nu_1\gamma_1\lambda_1$.
This equation together with Eq. (\ref{eq:hPerp}), where $h_\perp$ is given by Eq. 
(\ref{hperp}) determines $\psi$  and $u_{r\theta}$.
Eliminating $u_{r\theta}$, we find an equation for the steady state
orientation $\psi$
\begin{equation} \label{eq:Psistat}
h_\perp (4\eta+\gamma_1(\nu_1^2+1 -2\nu_1\cos 2\psi))=\gamma_1 \sin 2\psi(\nu_1\cos 2\psi -1)
\tilde\zeta\Delta \mu
\end{equation}
We now discuss special solutions to this equations as well as the stability
of asters and vortices obtained at thermal equilibrium for $\Delta\mu=0$.

\subsubsection{Spiral solutions for $\delta K=0$}

We first consider the simple case where there is no anisotropy of the elastic constants: $\delta K=0$ and $h_\perp=K \Delta \psi$. 
Eq. (\ref{eq:hPerp}) requires that spirals with constant $\psi_0$
and $h_\perp=0$ satisfy
$\cos 2\psi_0=1/\nu_1$.
This selection of angle, which depends on the parameter $\nu_1$ is a dynamic
phenomenon. The same angle is selected for the orientation of
nematic liquid crystals in shear flows. A steady state exists only if $\vert \nu_1 \vert \geq 1$.
The polarization angle takes  one of the two values
\begin{equation}
\psi_0 = \pm\frac{1}{2}\arccos\frac{1}{\nu_1}
\end{equation}
in the interval $[-\pi,\pi]$. 
With these values of $\psi_0$, we find
\begin{equation}
u_{r\theta} = \frac{\sin 2\psi_0}{4\eta +
\gamma_1\nu_1^2\sin^2 2\psi_0} \tilde\zeta \Delta\mu\quad.
\end{equation}
The velocity is ortho-radial, along the $\theta$ direction; it is obtained from (\ref{urt})
\begin{eqnarray}
v_\theta &=& 2 r \left (\int_0^r \frac{ u_{r\theta}}{r'} dr' +v_\theta^0\right) 
\end{eqnarray}
where $v_\theta^0$ is an integration constant. For a finite system with
radius $R$ and with the boundary condition that no motion occurs at the boundary,
\begin{equation}
v_\theta = 2 u_{r\theta}r\log\frac{r}{R}\quad.
\end{equation}

\subsubsection{Stability of asters and vortices}

Both asters with $\psi_0=0,\pi$ and
vortices $\psi_0=\pm \pi/2$ are solutions to Eq.(\ref{eq:Psistat}). In order to
understand under which conditions these solutions are stable, we perform a
linear stability analysis, writing $\psi(r)=\psi_0+\delta\psi$.
The time-dependence of $\psi$ in the liquid limit $\tau\rightarrow 0$ is given by
\begin{equation}
\gamma_1\frac{\partial\psi}{\partial t}=h_\perp-\gamma_1 (\nu_1\cos 2\psi-1)u_{r\theta}
\end{equation} 
Linearizing this equation and using Eq. (\ref{urt0}) we find for asters  with 
$\psi_0=0$ and $\delta K>0$ that
$\delta \psi(r)$ satisfies to linear order
\begin{equation}
\frac {\gamma_1}{1+\gamma_1 (\nu_1 -1)^2/(4\eta)} \frac{\partial}{\partial t} \delta\psi
= {\cal L}\delta \psi
\end{equation}
where the linear operator is given by
\begin{equation} \label{eq:deltaPsia}
{\cal L}= (K+\delta K)\left (
\frac{d^2 }{dr^2}+\frac{1}{r} \frac{d}{dr}
-\frac{n^2}{r^2}+k^2\right ),
\end{equation}
Here, we have defined $n^2=\delta K/ (K+\delta K)$ and $k^2=-\frac{2\tilde \zeta\Delta\mu(\nu_1-1)\gamma_1}{[4\eta +\gamma_1
(\nu_1-1)^2](K+\delta K)}$.
An aster solution becomes unstable, if the largest eigenvalue of
${\cal L}$ vanishes. The condition ${\cal L}\delta \psi=0$ is solved
by Bessel functions of order $n$, $\delta \psi\sim J_n(k r)$.
For a finite system with radius $R$ we find with boundary conditions
$\delta\psi(R)=0$ that $kR=z_n$, where $z_n$ denotes the first positive
root of $J_n(z)$. Therefore, the critical value for the instability of asters
is given by
\begin{equation} \label{eq:mucaster}
\tilde\zeta\Delta\mu_c^A =
-\left(\frac{z_{n}}{R}\right)^2\frac{4\eta +
\gamma_1 (\nu_1 -1)^2}{2\gamma_1
(\nu_1-1)} (K+\delta K)
\end{equation}

In the case of vortices with $\psi_0=\pm\pi/2$
and $\delta K<0$, we find 
\begin{eqnarray} \label{eq:deltaPsiv}
\frac {\gamma_1}{1+\gamma_1 (\nu_1 +1)^2/(4\eta)} \frac {\partial\delta \psi}{\partial t}
={\cal L}'\delta\psi \nonumber \\
{\cal L}'= K \left (
\frac{d^2 }{dr^2}+\frac{1}{r} \frac{d}{dr}
-\frac{m^2}{r^2}+q^2\right ),
\end{eqnarray}
where $m^2=-\delta K/K$ and $q^2=-\frac{2\tilde \zeta\Delta\mu(\nu_1+1)\gamma_1}{[4\eta +\gamma_1(\nu_1+1)^2]K}$.
Following the same analysis as for asters, the instability of vortices occurs for
\begin{equation} \label{eq:mucvortex}
\tilde\zeta\Delta\mu_c^V =
-\left(\frac{z_{m}}{R}\right)^2\frac{4\eta +
\gamma_1 (\nu_1 +1)^2}{2\gamma_1
(\nu_1+1)} K
\end{equation}

The resulting stability diagram in the
$(\tilde{\zeta}\Delta \mu,\delta K)$-plane is displayed
on Fig. 2. 
We have assumed on this diagram that $\nu_1\geq 1$. There are three regions 
on this diagram, a region with positive $\delta K$ and positive $\tilde\zeta \Delta\mu$
where the asters are stable, a region with negative $\delta K$ and positive $\tilde\zeta \Delta\mu$ 
where vortices are stable and a region with large negative $\tilde\zeta \Delta\mu$ where spiral 
defects are stable. Along the line $\delta K=0$, spirals are always stable as studied in 
the previous paragraph. We do not give here a general description of the spiral defects, as in the case where  $\delta K=0$, we expect that there is
a dynamic selection of the orientational angle and that the defect is rotating.

\subsection{Effects of friction with the substrate}

So far we have considered the point defect in a real  two-dimensional space. A real active 
gel interacts 
with both the surrounding solvent and the substrate.
We do not discuss the hydrodynamic interactions between the gel and the solvent here. We now consider a thin quasi-two dimensional 
gel layer and introduce a friction force at the substrate proportional to the local velocity. Taking into account the rotational symmetry, the friction force modifies 
the ortho-radial ($\theta$) component of the force balance that becomes

\begin{equation}
  \label{eq:tangstress}
 \frac{\partial \sigma_{r\theta}}{\partial r}+
\frac{2}{r}\sigma_{r\theta}=\xi v_{\theta}, 
\end{equation}

This equation, which is a balance between the friction force and the tangential
stress
can be integrated for a rotating spiral defect, yielding the following
expression for the tangential stress as a function of the velocity.

\begin{equation} \label{eq:sigmaq2d}
\sigma_{r\theta}=\frac{\xi}{r^2}\int_0^r dr'\; r'^2 v_\theta(r')
\end{equation}

In the absence of elastic constant anisotropy, $\delta K=0$, we can derive the velocity profile $v_\theta(r)$  in an
infinite system from Eq.~(\ref{eq:sigmaq2d}) as follows:
We look for a solution of Eq.~(\ref{eq:hPerp}) with $h_\perp=0$ and
$\psi=$const.. As in the absence of friction, the polarization angle can have one of the 4 possible values satisfying $\cos 2\psi_0=1/\nu_1$.
The velocity field is then obtained from Eq.~(\ref{eq:uRT}) that can be transformed 
into the following ordinary
differential equation for $v_\theta$:
\begin{equation}
 \frac{d}{dz}z^3\left(-\frac{v_\theta}{z^2}+\frac{1}{z}\frac{dv_\theta}{dz}\right)
=z^2 v_\theta+2\lambda_f \omega_0 z, 
\end{equation}
where $z=r/\lambda_f$ and the the friction length is given by
${\lambda_f =(4\eta+\gamma_1\nu_1\sin^2 2\psi_0)^{1/2}/(2\xi^{1/2})}$.
The velocity is then given by 
\begin{equation} \label{eq:vq2d}
v_\theta(r)= 2\omega_0 \lambda_f \left\{
K_1(\frac{r}{\lambda_f })-\frac{\lambda_f}{r}  \right\},
\end{equation}
where $K_1$ is the modified Bessel function of the second kind defined in
Ref. \cite{abra72},

At short distances ($r\ll \lambda_f$), the dissipation is dominated by the
viscosity $\eta$
and the velocity field is the same as in the absence of friction. At
large distances,
the dissipation is dominated by the  friction on the substrate $\xi$ and the
velocity $v_\theta$
decays to zero. The stability diagram of the defect in the presence of substrate 
friction is very similar to the diagram of Fig. ({\bf 3}) where the finite size $R$ would be replaced by the friction length $\lambda_f$.

\section{Discussion}

We have introduced in this manuscript  equations which describe the long wavelength and 
low frequency behavior of active gels. Although we have written the equations specifically in the case where 
the activity is 
due to motor proteins, they should apply, in their principles, to all gels in which a permanent 
source of dissipation is at work. Such gels define a new class of materials. For instance, 
a conventional "physical" polar gel, absorbing a high frequency ultrasonic wave should 
obey, in the low frequency, long wavelength limit, the set of equations proposed here. 
Our main motivation however, is the construction of a generic theory for 
characterizing quantitatively the properties of Eukaryotic biological gels. One could object 
that generalized hydrodynamic  theories involve a large number of parameters and are thus 
not very useful. Their merit is to involve the smallest number of parameters required for a 
relevant description of the systems under consideration and to describe in a unified way all long 
length scale and long time scale situations which are otherwise seemingly unconnected. The nature 
of most parameters is already well-known for  gel or polymer rheology  or for liquid crystal physics, and 
their measurement techniques can straightforwardly be transfered to active gels. 
This is transparent for translational,  orientational elastic moduli and viscosities. Coefficients 
linking shear flow and polar orientation are less familiar to the general public but are well known 
to liquid crystal physicists, and their measurements are not difficult a priori. There are  six bulk and 
two surface additional parameters, compared to a passive polar gel. Among the bulk quantities one is equivalent to the 
motor velocity on the actin filaments ($\lambda)$, and another one is the depolymerization rate $k_d$ of the gel in the 
bulk. Both quantities are directly accessible to experiment. Surface polymerization terms 
can be extracted independently from biochemical data and although they generate a 
new interesting physics (\cite{howa01,pros01}), 
they do not introduce uncertainties in the description.

Among the four remaining parameters, three ($\zeta$, $\bar\zeta$ and $\zeta'$
bear essentially the same physics (i.e: 
the activity implies either spontaneous motion or spontaneous stress), and the last one $\lambda_1$
measures the active rate of change of polarization.
Hence,  there are only two 
relevant additional parameter, namely $\lambda_1$ and 
$\zeta$ with respect to a passive polar gel. We show, in the two 
examples  developed here, that they change profoundly the behavior of gels: behaviors
which in the absence of energy input would be static, become dynamic. 
The structure of the relaxation modes
which would be entirely over-damped in the long wavelength limit, now can 
support propagative waves \cite{rama00,simh02}, spiral disclinations rotate permanently. Many other 
consequences have to be unraveled. Knowing  that it has taken more than ten years 
to investigate the properties of generalized hydrodynamical equations relevant to liquid 
crystals, it is likely that a similar number of years will be necessary in this case as well. 
Our bet is that it will help us understand the  complex behavior of the slow dynamics of 
the eukaryotic cytoskeleton in a robust way, not depending on the details of the involved 
proteins. 

 Our theory can be extended in several ways. First we have ignored the permeation 
 process of the bulk fluid through the gel. This is legitimate in the long time limit in 
 most geometries but not all. The inclusion of  permeation in our equations would 
 be straightforward. We have also used the simplest visco-elastic gel theory: 
 experiments performed on cells suggest the existence of scale invariant visco-elastic 
 behavior which could also be included in the theory \cite{fabr01}. At last we have assumed that 
 the driving force was "small" and constant both in space and time. The spatial and 
 temporal invariance make sense in a cell since the $ATP$ production centers seem to 
 be abundant and evenly distributed. In vitro experiments may mimic these conditions 
 on a limited time scale, yet large compared to most phenomena of interest. 
 Furthermore there is no a priori difficulty in introducing $ATP$ production and 
 consumption in the equations. More severe is the last limitation: our equations are valid for 
 $\Delta\mu$ small compared  to thermal energy, whereas it is of the order of ten time that 
 value in real life. The extension of our theory to large $\Delta\mu$ 
 is a real challenge and will probably require some feedback from experiments. A 
 brute force expansion in  higher powers of forces and fluxes is just totally 
 impractical and would leave out the known and probably leading exponential 
 dependences of polymerization/depolymerization rates on stress. Other 
 challenging questions concern the nature of noise which will pick up strong 
 out of equilibrium contributions, and possible dynamical transitions for instance 
 in the motor behavior\cite{juli97}. For these reasons, we think that is it wise to start with the 
 well controlled approach that we propose here and to develop the corresponding experiments. 
 This will help us familiarize with this new physics before getting to regions of phase space 
 were we lack guiding principles.

An important application of active gels is the problem of cell locomotion on a substrate \cite{sperm}.
In a forthcoming publication, we will discuss how the interplay of polymerization
and activity can induce the motion of a thin layer of an active gel on a solid surface.
The physics of this gel motion coupled to the properties of bio-adhesion  provides a
basis for an understanding of the locomotion of cells such as keratocytes \cite{verk98}.

\begin{appendix}

\section{Cylindrical coordinates}
\label{sec:app1}
In polar coordinates,
the rate of strain tensor for a rotationally symmetric velocity profile is given by
\begin{eqnarray}
u_{rr} & = & \frac{d}{dr}v_r\\
u_{\theta\theta} & = & \frac{v_r}{r}\\
u_{r\theta} & = &
\frac{r}{2}\frac{d}{dr}\left(\frac{v_\theta}{r}\right)\label{urt}\quad.
\end{eqnarray}
Furthermore,
\begin{equation}
\omega_{r\theta} = \frac{1}{2r}\frac{d}{dr}(rv_\theta)\quad.
\end{equation}

If the system is invariant by rotation, the local force balance in cylindrical coordinates is written as
\begin{eqnarray}
\frac{1}{r}\frac{d}{dr}\left(r\sigma^{\rm tot}_{rr}\right) -
\frac{1}{r}\sigma^{\rm tot}_{\theta\theta} & = & \frac{d\Pi}{dr}\label{srrtt}\\
\frac{1}{r^2}\frac{d}{dr}\left(r^2\sigma^{\rm tot}_{r\theta}\right) & = & 0\label{srt}\quad,
\end{eqnarray}
In an incompressible system, the pressure $\Pi$ is a Lagrange multiplier necessary to impose the
incompressibility constraint.

\section{Viscoelastic polar gel}
The form of Equations (\ref{sigmadab}), (\ref{pda}), (\ref{sigmar}) and (\ref{Pr}) can be
understood by discussing the general dynamics of the polarization field $p_\alpha$
in frequency space. As a function of frequency, we write to linear order
 \begin{equation}
i\omega \tilde p_\alpha=\frac{\tilde h_\alpha}{\gamma_1}+\lambda_1\tilde p_\alpha \Delta\mu
-\frac{1}{1+i\omega\tau}(\nu_1 p_\beta\tilde u_{\alpha\beta} +\bar \nu_1 p_\alpha\tilde u_{\beta\beta}) ...
\label{pomega} 
\end{equation}
where the tilde denotes a Fourier amplitude at frequency $\omega$.
The first two terms of the right hand side 
correspond to a straightforward linear response theory of a
ferroelectric liquid crystal. Similarly, the last term provides for $\omega=0$ the
reactive linear coupling term allowed by symmetry. For finite frequency furthermore,
this term
takes into account that the gel is
elastic on short time scales or large $\omega$. This can be seen as follows:
in the case of an elastic gel, the free energy $F$ of the polarization field
contains elastic terms of the form
\begin{equation}
F_{\rm el}= \frac{\gamma_1\nu_1}{2\tau} \int d^dx p_\alpha \epsilon_{\alpha\beta}p_\beta
\end{equation}
where $\epsilon_{\alpha\beta}$ is the strain tensor characterizing deformations of the gel.
This leads to a modification of $h_\alpha$ by $h_\alpha^{\rm el}=-\delta F_{\rm el}/\delta p_\alpha$.
Since in frequency space, $\tilde u_{\alpha\beta}=i\omega\tilde \epsilon_{\alpha\beta}$,
Eq. (\ref{pomega}) does describe the correct elastic limit for large $\omega$
Separating Eq. (\ref{pomega}) in reactive and dissipative parts, using the opposite
signatures with respect to time reversal (or equivalently $\omega\rightarrow -\omega$)
allows us to identify $P_\alpha^d$ and $P_\alpha^r$.
Using $1/(1+i\omega\tau)=1-i\omega\tau/(1+\omega^2\tau^2)$
and replacing $i\omega$ by $D/Dt$,
 we can write the reactive and dissipative parts of $P_\alpha$ as given by
Eqns.  (\ref{pda})and (\ref{Pr}).
Symmetry relations now impose the correct equations for $\sigma^d_{\alpha\beta}$ and
$\sigma^r_{\alpha\beta}$ as given by Eqns. (\ref{sigmadab}) and (\ref{sigmar}).
\end{appendix}

\newpage
\begin{figure}
\scalebox{0.8}{
\includegraphics{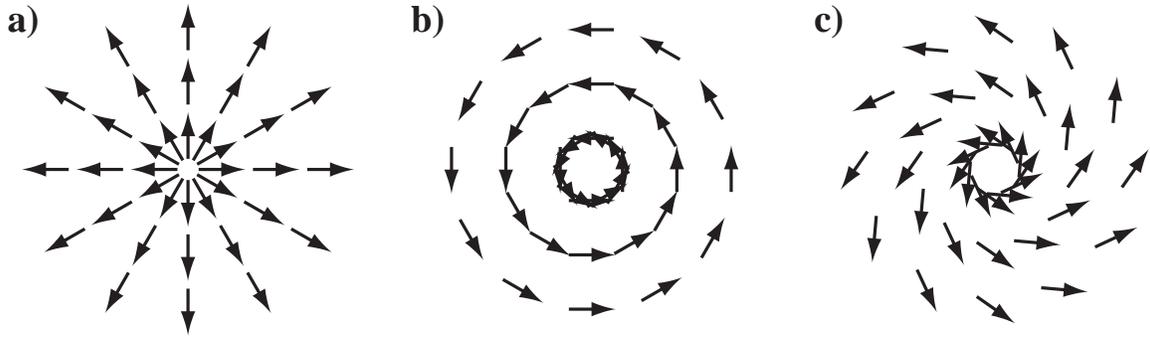}}
\caption{Schematic representation of the geometry of point defects of 
topological charge one in
two dimensions. Displayed are the orientations of polarization vectors 
as arrows for (a) aster, (b) vortex and (c) spiral.}
\end{figure}
\begin{figure}
\includegraphics{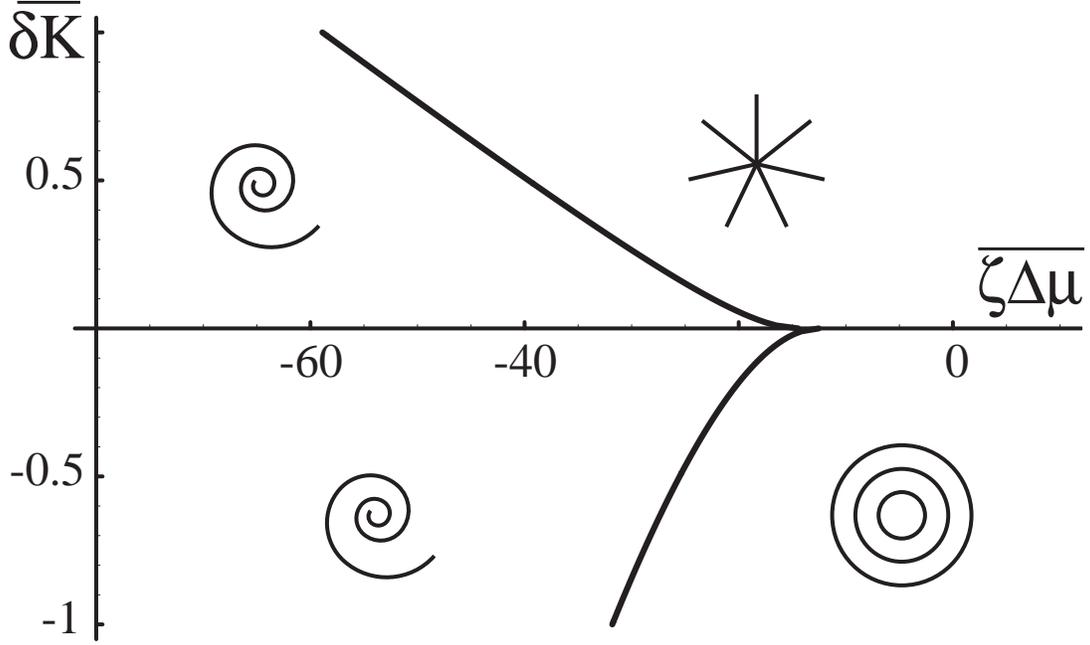}
\caption{Stability diagram of asters, vortices and spirals which are 
topological defects in an active gel of polar filaments. Asters are stable 
for $\delta K>0$ in the region with  actively generated stresses 
$\tilde\zeta\Delta\mu$ larger than a critical value $\tilde\zeta\Delta\mu^A_c$. 
This critical value is negative, corresponding to contractile stresses in the
gel. Vortices are stable for $\delta K<0$ and 
$\tilde\zeta\Delta\mu>\tilde\zeta\Delta\mu^V_c$. For other parameter values, 
rotating spirals occur via a symmetry breaking dynamic instability. Here, 
$\overline{\delta K} = \delta K/K$ is a dimensionless ratio of two elastic
moduli and $\overline{\zeta\Delta\mu}=R^2\tilde\zeta\Delta\mu/K$ a 
dimensionless measure of active stresses. Note that both rotation senses
occur with equal probability on symmetry grounds. The diagram was evaluated for the 
choice $\eta/\gamma_1=1$ and $\nu_1=2$ of Onsager coefficients of the system.}
\end{figure}

\end{document}